\documentclass{IEEEtran}
\usepackage{amsmath,amsfonts}
\usepackage{algorithmic}
\usepackage{array}
\usepackage[caption=false,font=normalsize,labelfont=sf,textfont=sf]{subfig}
\usepackage{textcomp}
\usepackage{stfloats}
\usepackage{url}
\usepackage{verbatim}
\usepackage{graphicx}
\usepackage{xcolor} 
\usepackage{array}
\usepackage{graphicx}
\DeclareGraphicsExtensions{.pdf,.png}
\usepackage{tikz} 
\usepackage{comment}

\usepackage{algorithm2e}
\usepackage{amsfonts}
\RestyleAlgo{ruled}
\usepackage{fmtcount} 
\usepackage{array,multirow}

\usepackage{acronym} 

\newcolumntype{C}[1]{>{\centering\arraybackslash}m{#1}}

\usepackage{booktabs}
\usepackage{graphicx}
\def\BibTeX{{\rm B\kern-.05em{\sc i\kern-.025em b}\kern-.08em
    T\kern-.1667em\lower.7ex\hbox{E}\kern-.125emX}}
\usepackage{balance}
\begin{document}
\title{Graph Neural Network Framework for Security Assessment Informed by Topological Measures}
\author{Mojtaba Dezvarei, Kevin Tomsovic, Jinyuan Stella Sun, Seddik M. Djouadi
\thanks{This work was partially supported by NSF grant CNS-2038922.} 
\thanks{Mojtaba Dezvarei, Kevin Tomsovic, Jinyuan Stella Sun, and Seddik M. Djouadi
are with the Department of Electrical Engineering and Computer Science, The University of Tennessee, Knoxville, TN 37996 USA. (e-mail: mdezvare@utk.edu; tomsovic@utk.edu; jysun@utk.edu; mdjouadi@utk.edu)}
}


\maketitle

\begin{abstract}
In the power system, security assessment (SA) plays a pivotal role in determining the safe operation in a normal situation and some contingencies scenarios. Electrical variables as input variables of the model are mainly considered to indicate the power system operation as secure or insecure, according to the reliability criteria for contingency scenarios. In this approach, the features are in grid format data, where the relation between features and any knowledge of network topology is absent. Moreover, the traditional and common models, such as neural networks (NN), are not applicable if the input variables are in the graph format structure. Therefore, this paper examines the security analysis in the graph neural network (GNN) framework such that the GNN model incorporates the network connection and node's neighbors' influence for the assessment. Here the input features are separated graphs representing different network conditions in electrical and structural statuses. Topological characteristics defined by network centrality measures are added in the feature vector representing the structural properties of the network. The proposed model is simulated in the IEEE 118-Bus system for the voltage static security assessment (SSA). The performance indices validate the efficiency of the GNN-based model compared to the traditional NN model denoting that the information enclosed in graph data boosts the classifier performance since the GNN model benefits the neighbors' features. Moreover, outperforming of GNN-based model is determined when robustness and sensitivity analyzes are carried out. The proposed method is not limited to a specific task and can be extended for other security assessments with different critical variables, such as dynamic analysis and frequency criteria, respectively. 
\end{abstract}


\section{Introduction}
\IEEEPARstart{S}{ecurity} itself means being free from risk and threat. In this context, power system can not be considered secure since it always anticipates the occurrence of disruption. Also, the possibility of an incident is increasing since the power system is being utilized inverter-based resources (IBR) such as wind turbines and solar panels \cite{negnevitsky2019high}. So, security in the power system measures the risk or danger of disruption of the continuously operating system. In practice, that is the ability of the system to withstand sudden disturbances with minimum
disruption to its performance. 

From an operational point of view, the power system is secured if important electrical variables keep within an acceptable level, such as bus voltage magnitudes and angles, frequency, and power flows in response to disturbances (\textit{contingencies}) like an electric short circuit, change of transmission system configurations due to faults or sudden load increase. Security Assessment (SA) is an analysis performed to determine whether and to what extent a power system is safe from serious interference in its operation. The SA carry outs to meet the operation requirement in two manifests: $(1)$ surviving the ensuing transient and moving into an acceptable steady-state condition, and $(2)$ in this new steady-state condition, all components are operating within established limits \cite{sittithumwat2002dynamic}. Followed by this time framework, Static Security Assessment (SSA) evaluates the steady-state response and Dynamic Security Assessment (DSA) analyzes the transient response. The SSA is the topic of concern for this paper as utilities companies mainly take this into account for planning and operation purposes. The acceptable level of variables such as transient voltage magnitude dip and steady-steady violation or frequency excursion is determined by reliability criteria provided by, for instance, Western Electricity Coordinating Council (WECC) or North American Electric Reliability Corporation (NERC) standards.

The SA can also target to determine the frequency-related violation or voltage violation for both dynamic or static analyses. The former is becoming more challenging in power system operation and planning, especially with the increasing penetration level of IBR, which
brings about insufficient inertial and primary frequency responses \cite{inertia1,9494275}. The authors in \cite{steinkohl2019frequency} consider the frequency security analysis to provide an appropriate frequency control scheme when power systems are utilized the wind energy or frequency security index. This determines the frequency security based on all aspects of the frequency profile presented to specify the relative distance from the security margin \cite{arias2015ieee}. The significance of voltage violation is apparent as the reason for several large blackouts is due to inadequate reactive power supply. For this purpose, the SA may perform to either estimate the distance to the nose point of $P \text{-}V$ and/or $V\text{-} Q$ curves as a margin \cite{Taylor}, or classify multiple operating conditions for voltage stability assessment \cite{vanfretti2020decision}.

The SA is also categorized for security classification or security margin estimation purposes. Classification determines whether the power system is secure or insecure with regard to the threshold, whereas in security margin, the distance (margin) to the insecure condition, the violation threshold, is computed. For instance, an online voltage security assessment practice to prevent a large-scale blackout and an estimation loading margin regarding transient frequency criteria capture classification and estimation tasks, respectively \cite{sittithumwat2005dynamic,negnevitsky2015random}. This paper focuses on classifying static voltage in the SA, followed by contingency scenarios.

Besides the traditional methods for the SA, such as lookup tables and nomograms, in which the operator decision was essential, the automated SA mechanism uses a model to determine the security statutes regarding the value of system variables and the measurement of so-called features. Thanks to data availability by sensors and/or parameter estimation, extensive usage of artificial intelligence (AI) methods are found in the literature for the SA. The applications of NN can be to evaluate the system security by screening the credible contingencies with loading condition and the probable contingencies as the input \cite{sunitha2013online}, and to estimate loadability margin concerning frequency deviation with preventive
control \cite{sittithumwat2002dynamic}.
A real-time SA to increase awareness about plausible future insecurity is applied using the Decision Tree (DT) form collection of Phase Measurement Units (PMUs) data \cite{mukherjee2021real}. An attempt to classify whether the power system can tolerate an $(N-1)$-fault during different conditions is analyzed via Support Vector Machines (SVM) with the practice of Principal Component Analysis (PCA) for dimensional reduction of feature space \cite{anderssonpower}. In addition to NN, the promising results of deep learning (DL) frameworks for image and speech recognition caused an emerging usage of DL for SA purposes to capture immense amounts of data and deliver valuable information. As a typical network model, Convolutional Neural Network (CNN) can be exploited for power system transient stability assessment, instability mode prediction, and small-signal stability \cite{shi2020convolutional,arteaga2019deep}.

All above models are associated with grid structure data for input features, meaning that a fixed size of grid data assuming that instances are independent, is fed to the models. For example, even if a graph represents a grid data format like an image, it has a banded structure in its adjacency matrix since all nodes are formed in a grid. This is no longer valid for graph data as the number of neighbors to each node is variable, and  difference in size and shape of within graph dataset can not handle using resizing or crop operation in images. As a unique non-Euclidean data structure and the need of permutation invariant for for machine learning model due of graph isomorphism, GNN model is introduced. Graph analysis focuses on node classification, link prediction, and clustering tasks. Indeed, GNN models are DL-based methods performing on the graph domain.

Due to the promising results of GNN in social science, natural science, protein-protein interaction networks, and knowledge graphs, broad usage of GNN models can be noticed in the literature \cite{wu2020comprehensive}. However, the application of GNN in power systems is not vast compared to other domains, and usage of the GNN models in power systems is discussed in \cite{liao2021review}. For example, the GNN model using a power flow solution exploited for  
predicting the electricity market prices addresses scalability and adaptivity challenges of existing end-to-end optimal power flow (OPF) learning results \cite{liu2021graph}. In security concerns, \cite{huang2020recurrent} provides a scheme combining GNN and recurrent neural networks for stability classification and critical generator identification in transient assessment. A graph convolutional network (GCN) framework can be applied for fault location identification in a distribution network, indicating GCN model robustness to limited bus measurements and outperforming other machine learning models \cite{chen2019fault}. As there is a lack of a GNN model for the SSA, this paper seeks to form a GNN framework for voltage SSA.  

Regardless of which model is used for the SA task, electrical variables such as active/reactive power line flow, bus voltage angle, and magnitudes, active and reactive power of each bus load are mainly considered for input features. This feature space lacks the topological knowledge of the power system, as one may represent the power grid as a graph where buses and lines are illustrated as nodes and edges, respectively. The large scalability of the power system motivates researchers to model it as a graph to study the system vulnerability in the topological context using centrality measures. These measures may indicate the most salient part against random failures and directed attacks \cite{wang2010electrical,hines2010topological}. In this context, the power system characteristics coming from topological information may assist in analyzing the impact of network structure changes to enhance the model performance and ensure robustness for the SA. The centrality measures can be easily computed using the network topology processor's information implemented in Energy Management System (EMS). To the authors' best knowledge, there is no work in the literature to examine electrical and topological-related variables in the SA framework. Therefore, this paper proposes a method for voltage SSA based on the GNN model that combines electrical variables obtained by power flow solution and topological parameters defined by grid centrality measures.

With regard to the importance of SSA in power system operation and planning and considering the increment of uncertainty and incident in the power system, this paper is to deliver a resilient framework for static voltage security assessment. The proposed framework is validated in IEE 118-Bus, in which the results indicate that the GNN-based SA model outperforms the traditional NN model. Through the model robustness and sensitivity investigation, the GNN-based model also presents better performance metrics revealing that the proposed model is more capable of capturing uncertainty and obtaining promising output. The main contributions of this work are:

\begin{enumerate}
\item The SA schemes lack the knowledge of topological changes occurring during contingency scenarios or unplanned incidents and change the power grid structure. This paper considers the power grid as a graph to add topological information to the electrical features space. All structural changes can then be observed and measured using graph centrality measures. The new feature space is more informed about electrical and structural variables.

\item The common practice in the SA is to use the grid format data as independent features. In this fashion, the connection information in the graph representation is missed. After presenting the power grid as a graph with both electrical and topological features, this paper delivers the SA model based on the GNN model, where each sample is a graph of the power grid after an incident where features are embedded in each node. Security classification is then transferred to graph-level classification using the GNN model. Indeed,  a graph dataset encompassing both electrical and structural features per bus (node) for each sample is used to classify the graph as secure or insecure with defined security criteria. The advantage of the GNN model is to benefit the local information of each node where during model training, the node features are more knowledgeable due to shared and updating information of neighbors nodes. 
\item The proposed approach is pretty straightforward to follow and implement. In addition to the electrical variables generator by power flow results, it only needs to have a graph measures generator for centrality measures. Furthermore, the proposed SA GNN-based is constructed as a comprehensive framework to examine the DSA in frequency, voltage violation, etc. Moreover, as the GNN is an active research area, researchers have presented efficient techniques for large-scale graphs, indicating no issue with the scalability of the proposed SA model for real power systems.

\end{enumerate}
The rest of the paper is organized as follows. The preliminaries definition are provided in Section \ref{Sec1}. The problem formulation for the SA is expressed in Section \ref{Sec2}. The proposed GNN scheme with regard to topological measures is discussed in Section \ref{Sec3}. The simulation procedure and the results of the proposed method on the IEEE 118-bus system are indicated in Section \ref{Sec4}, followed by model robustness and sensitivity analyses in Section \ref{Sec5}. Finally, a discussion about the suggested method followed by a conclusion is stated in Section \ref{Sec6}.

\section{Preliminaries}\label{Sec1}
A graph $G$ is defined as $G=(V,E)$, where set $V$ is the set of vertices (nodes) and $E$ is the set of edges. Here, $V$ and $E$ are always finite. An edge ${x,y}$ is said to join the vertices $x$ and $y$ and is denoted by $xy$. A \textit{directed graph} is a connected one where all the edges are directed from one vertex to another. In contrast, a graph where the edges are bidirectional is called an \textit{undirected graph}.
The \textit{Adjacency matrix} of the graph $G = (V, E)$ is an $n\times n$ matrix $A = (a_{ij})$, where $n$ is the number of vertices in $G$, $V=\{v_{1},\dots, v_{n}\}$ and $a_{ij} =$ number of edges between $v_{i}$ and
$v_{j}$. When $a_{ij} = 0$, $(v_{i}, v_{j})$ is not an edge in $G$. The matrix $A$ of an undirected graph is symmetric, i.e. $A^T = A$. In the case of a directed graph, the same definition remains while the matrix $A$ is no more symmetric and depends on the edges direction. The \textit{Laplacian matrix} or Kirchhoff matrix of a graph carries the same information as the adjacency matrix but has different valuable and vital properties, many relating to its spectrum. The laplacian matrix is defined as $L = D - A$, where $D$ is a diagonal matrix indicating the node degree matrix. 
\section{Problem statement}\label{Sec2}
In this section, the static voltage security analysis is discussed to determine the security status of the power system. Many sources make power systems vulnerable, such as natural calamities, complement failure, fault, internal or external intrusion, human error, etc. In this case, the power system should be able to continue servicing in case of unpredicted contingency. If any vulnerability sources interrupt services, like an outage or blackout occurrence due to cascading failure, the system is insecure (vulnerable). In static security assessment, post-contingency time framework, regardless of transient behavior, is taken into account where the system reaches out to a new steady state operating points. If the new operating points meet the defined system limitation and reliability criteria, the system is said to be statically secure. In this fashion, a fast and reliable solution is necessary to assess the security of numerous operating strategies to reduce the risk of catastrophic incidents. This attempt is involved due to vast sources of vulnerabilities, the large scale of the power system with nonlinear behavior, different operating and operational strategies, changes in topologies, and the computation burden. Therefore, a classification model is an effective approach that can deal with with the difficulties. Classifiers' merits are that they can be developed offline, current and future operating states can be quickly assessed, and classifying a new steady state power system condition into a secure or insecure class is trivial and does not need protracted computations of an analytical solution.

In the SSA framework, the security variables could be bus voltages or line flows indicating the thermal limits. This paper considers bus voltage as a security variable; however, the proposed scheme can be applied to other variables and categories, such as dynamic assessment. In the SSA, following a contingency, each bus voltage value is analyzed at post-contingency in which the transient response has died down. As a result, the steady state of voltage (power flow solution) should not violate the range defined by operating limits. It is worth noting there is no updated guideline for the voltage violation range at the steady-state analysis when the power grid is utilized by the IBRs. However, its impact on the transient response of voltage trajectory for reliability criteria is discussed in \cite{guideline2018bps}. In this regard, the problem statement is straightforward: \textit{following a contingency scenario, we are seeking to define a model that classifies the post-contingency condition of the power system (steady state) based on bus voltages according to reliability criteria}. Therefore, the SA model is to classify secure and insecure conditions and notifies the operators to steer the system away from the insecure state in adequate time.

To address the defined problem, this paper proposes applying the GNN framework for the SA. As the power grid may be analyzed in the graph representation, the problem here is formulated as \textit{graph-level classification}, and a graph including its information is assessed for security purposes.  The details of the GNN-based model are described in Section \ref{Sec3}.  

Graph input may also contain hidden information in its structural property, and shared local node information may enhance the model performance. Traditional and most applied models work with grid format input features to label the input dataset as secure/insecure for binary classification (or multi-class SSA \cite{kalyani2010classification}), where input features are independent, and there is no information between each feature. To benefit the edge connectivity of the power grid and append the local node information from neighbors to boost model performance, the GNN framework is developed. In this framework, the status of the power grid at steady state condition after contingency is considered as one sample instead of independent variables such as nodal voltage and line flow. For instance, the number of variables for one input of the non-GNN model could be the number of chosen features multiplied by the number of buses. In contrast, the GNN model only takes one graph as input, and all information is embedded in nodes or edges.
\section{The Proposed SA model}\label{Sec3}
\begin{figure}[ht]
    \centering
    \includegraphics[scale=0.48]{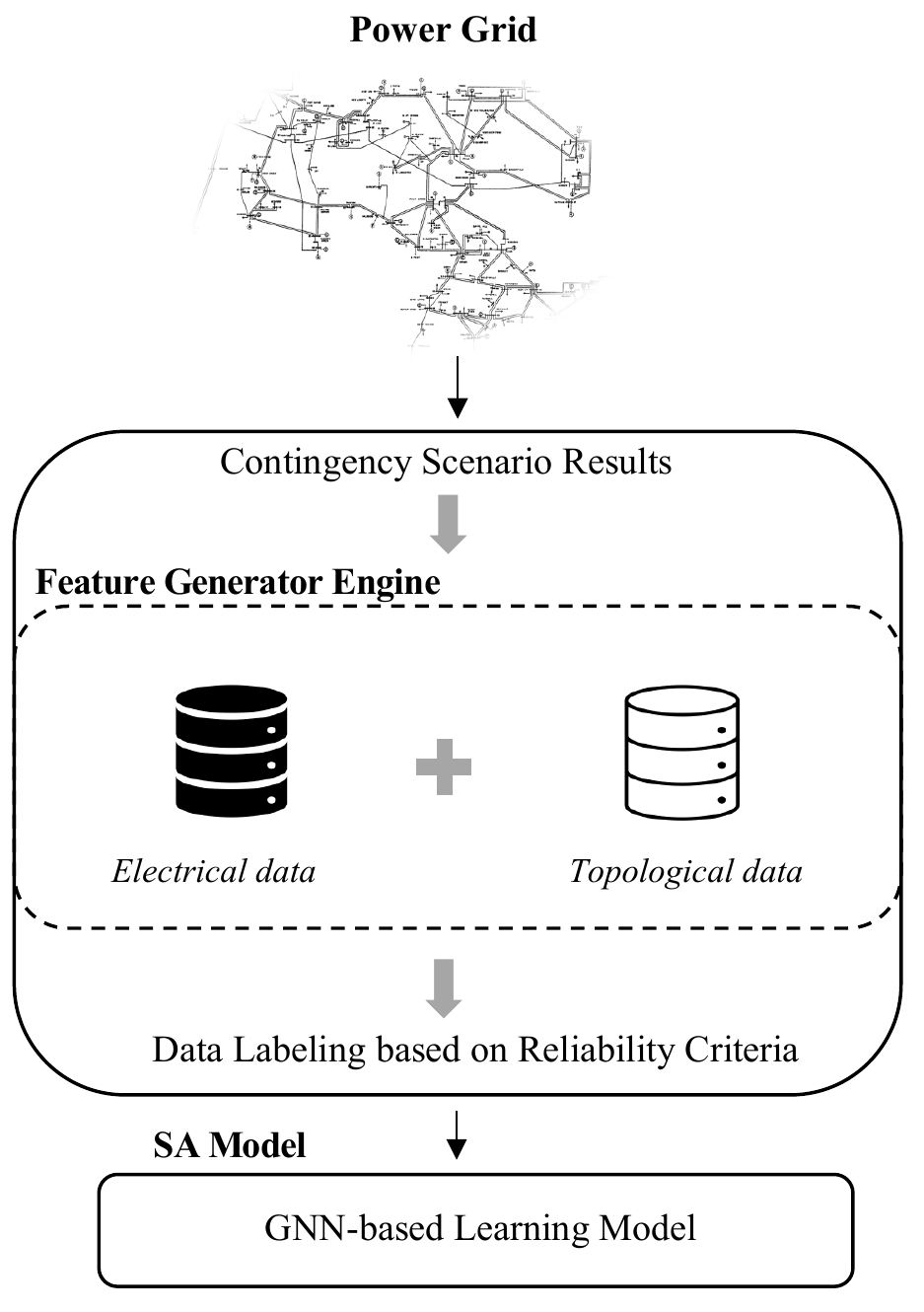}
    \caption{Procedure of the GNN-based SA model.}
    \label{fig3}
\end{figure}

Two main modules can explain the proposed procedure for the SA model as \textit{A. Feature Generator Engine} to create input features, and \textit{B. the GNN learning model} for the SA model shown in Fig. \ref{fig3}.
\subsection{Feature Generator Engine}
\subsubsection{Electrical Variables}
Electrical features are selected based on engineering knowledge associated with the SA problem and statistical correlation coefficients to ensure variables fluctuate in relation to each other and eliminate redundancy. The typical electrical variables include line active and reactive power, voltage magnitude, and angle. Considering voltage SSA problem, the following are chosen as electrical features in the input dataset: \begin{itemize}
     \item The voltage magnitude of each bus, \(V_{mag}\)
     \item The active and reactive of each bus, \(P,Q\)
 \end{itemize}{}
As the power grid provides numerous measurements as an extensive dataset, the SA may involve the curse of dimensionality. As only three leading electrical variables are considered in the procedure, the dimensional reduction (DR) approach can not be relevant. For numerous variables, the SA may involve the curse of dimensionality, which techniques such as PCA and fisher discrimination can be applied to identify the most significant and valuable subset of features for accurate classification \cite{anderssonpower, niazi2004power,jensen2001power}.
\subsubsection{Topological Variables} \label{top_cent}
Power grids have grown organically over the years in a random way to achieve economic benefits and safety, leading to a widely distributed grid with many connections between generation units and substations. Therefore, the complexity of the links and being a large-scale system lead researchers to study power grids in the context of graph representation using
statistical tools for vulnerability studies \cite{chu2017complex}. Table 5 in \cite{pagani2013power} reviews various resilience analysis and improvement studies in graph context. This motivates us to investigate the structural properties of the power grid and append topological features into feature space.

In addition to knowing the number of nodes and edges in a graph, it is worth learning the network's characteristics to indicate the important part of a network. The metrics so-called \textit{graph centrality} generally measure a unit's prominence; in different substantive settings, i.e., identify the most critical nodes in a graph given its topology with the various definitions of importance. Many various centrality measures have been proposed over the years \cite{latora2017complex}; in this paper, the most applicable measures in the power grid are discussed as follows to state as a new feature for each bus.

\textit{Degree centrality} ($C_{d}$) is a local measure and the simplest centrality measure. It implies that the nodes with a higher degree $deg(v)$ i.e., connected edges, are more solid, and the normalized degree centrality is defined as
\begin{equation*}
    C_{d}(v)= \frac{deg(v)}{n-1}=\frac{L(v,v)}{n-1}
\end{equation*}

\textit{Clustering Coefficient} ($C_{c}$) is a measure of degree to which nodes in a graph tend to cluster together. For each node $i$, it is the number of edges between neighbors of a node, divided by the total number of possible edges between those neighbors $C_{c_i}=2 e_{i}/{ k_{i}(k_{i}-1)}$ where $k_{i}$, $e_{i}$ are the number of he number of neighbors and connected edges between them, respectively.

\textit{Betweenness centrality} ($C_{b}$) consists both of a node and an edge. The node betweenness as most widely used reflects the influence of a node over the flow of information between other nodes, especially in cases where information flows over a network primarily follows the shortest available path. The node betweenness centrality is defined as 
\begin{equation*}
    C_{b}(v)= \frac{\sum\limits_{s\neq v \neq t \in V}^{}{\sigma_{st}(v)/\sigma_{st}}}
    {(n-1)(n-2)/2} 
\end{equation*}
where $\sigma_{st}$ is the number of shortest paths from $s$ to $t$ and $\sigma_{st}(v)$ is the total number from the mentioned paths that pass through vertex $v$. 

\textit{Closeness centrality} ($C_{k}$) is a way of detecting nodes that can spread information very efficiently through a graph. That is, a node is vital if it has a short distance from many other nodes and defined as
\begin{equation*}
    C_{k}(v)= \frac{\sum\limits_{t\in V\backslash v}{d_{G}(v,t)}}{n-1}
\end{equation*}
where $d_{G}(v,t)$ is shortest path length between vertices $v$ and $t$. This measures how far away a node is from the rest of the network instead of its closeness. Therefore, some researchers define closeness to be its reciprocal. 

The general concept of the shortest path in a graph is not an appropriate metric for the power grid. The shortest path (geodesic path) between two nodes in a graph is a path with the minimum number of edges (or minimum sum of edge weights for a weighted graph). This definition should be modified to cope with power grid characteristics since the electrical flow finds a path with minimum impedance value. So, \textit{electrical distance} $d_{Z}$ defined by \cite{wang2010electrical} is considered for computing shortest path distance based on line impedance as:
\begin{equation*}
    d_{Z}(v,t)= \| \sum_{(i,j) \in E \cap \ path(v,t)} Z_{pr}(i,j) \|
\end{equation*}
where $Z_{pr}(i,j)$ is the line impedance of the link $(i,j)$.

\begin{figure}[ht]
    \centering
    \includegraphics[scale=.57]{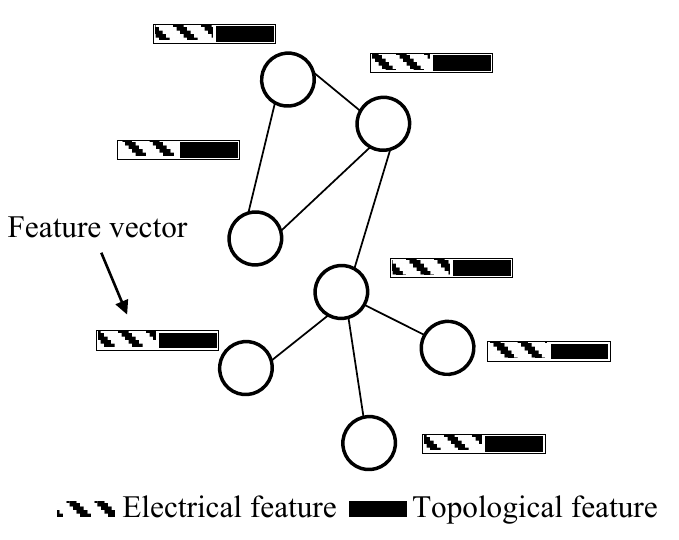}
    \caption{Embedding Feature Vector}
    \label{feature_vt}
\end{figure}

The measures obtain the influential nodes over the graph from a distinct aspect of view. Therefore,  the topological measures may aim to fill the possible gap between measures. Hence, beyond electrical variables, the above measures indicating the topological importance of nodes, are added into features for voltage SSA as two different database shown in Fig. \ref{fig3}. In other words, electrical and topological are embedded in each node as feature vector
\begin{align*}
\centering
[V_{mag}^{i}, P^{i}, Q^{i}, C_{d}^{i}, C_{c}^{i}, C_{b}^{i}, C_{k}^{i}],\quad \forall i\in \textit{Bus set} 
\end{align*}
The representation of feature vector is illustrated in Fig. \ref{feature_vt}, in which the graph input data with node embedding futures is deployed for the learning task.

\begin{figure*}[b!]
    \centering
    \includegraphics[scale=.54]{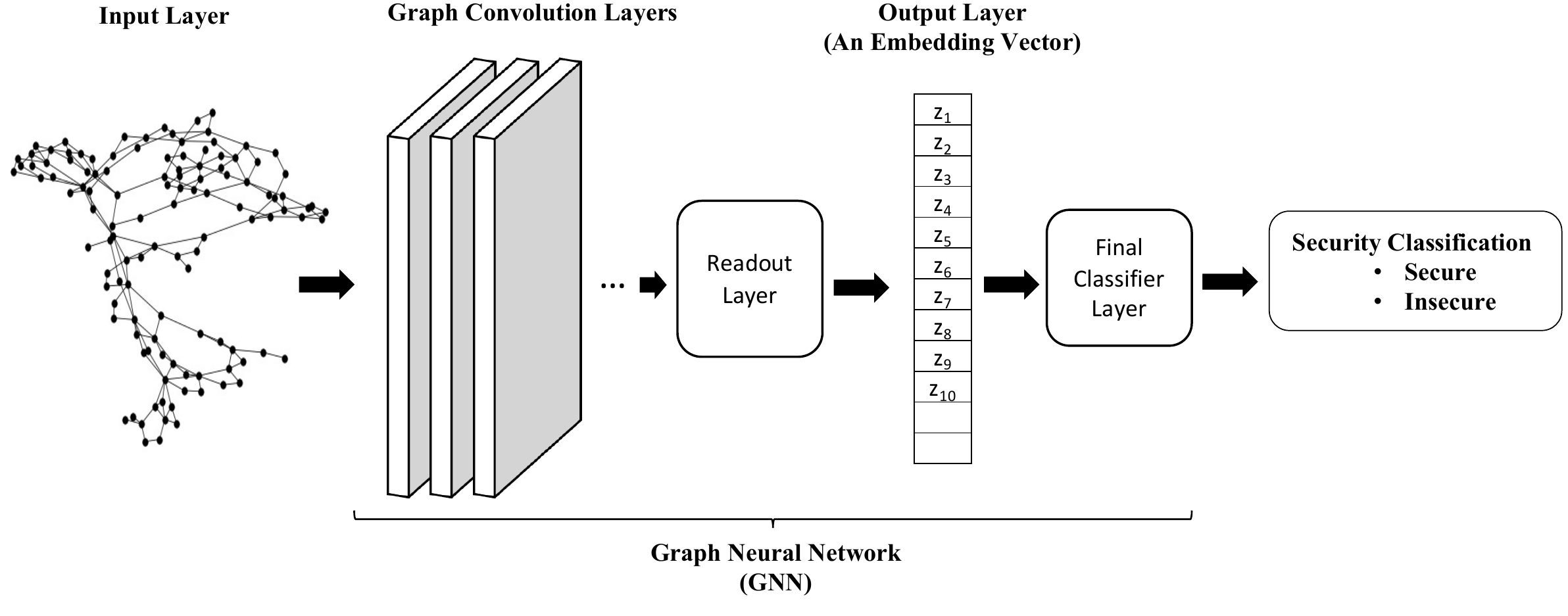}
    \caption{General GNN Framework for Classification Problem.}
    \label{fig1}
\end{figure*}

\subsection{Learning Model: Graph Neural Network}
In this paper, Graph Neural Network (GNN) framework is chosen as a learning model for voltage SSA. GNN is a type of deep neural network that is suitable for analyzing graph-structured data. In addition to the deficiency of Convolution NNs (CNNs) and Recurrent NNs (RNNs) with graph-structured data (well-defined only for grid-structured data and only over sequences, respectively, like images and texts), these models suffer the variation of size and shape of input as typically that except a fixed size of the input. Moreover, the CNN model is sensitive to input permutations, such as rotating an input picture. To address these challenges, GNN models are proposed to work with graph-structured data as they can handle the changes in shape and size of inputs and are permutation invariant. The overall framework of the GNN model for graph-level classification is shown as in Fig. \ref{fig1}.

\begin{figure}[ht]
    \centering
    \includegraphics[scale=.7]{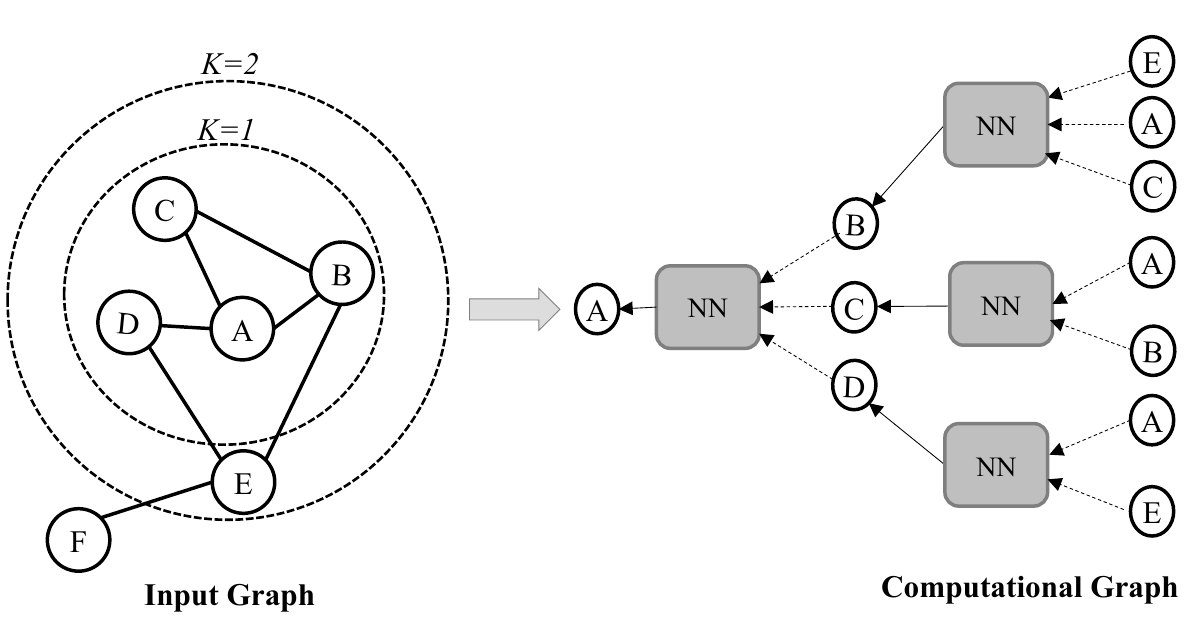}
    \caption{Computational graph of node A: aggregation of information from two hops away}
    \label{fig2}
\end{figure}

The main idea here is to generate representations of nodes that include the information on the graph's structure and any feature information it might have \cite{hamilton2020graph}. The procedure of GNN is encapsulated in \textit{neural message passing} in which the feature vector of the node is exchanged between nodes and updated using neural networks. There are two main components in GNNs:
\begin{itemize}
    \item \textit{Aggregate operator} $\mathcal{G}$: permutation invariant function to its neighbors to generate the aggregated node feature.
    \item \textit{Update operator} $\mathcal{F}$: combining the message from previous aggregated node feature to generate updated node embedding. 
\end{itemize}
$\mathcal{G}$ and $\mathcal{F}$ can be any arbitrary differentiable functions (i.e., neural networks), where $\mathcal{G}$ has to be permutation invariant operator.
The general procedure of aggregation and updates for a sample graph is shown in Fig \ref{fig2}. The node neighbors form a computational graph to aggregate and update information, such as from node A's local graph neighbors (i.e., B, C, and D). The messages coming from these neighbors are also based on information aggregated from their respective neighbors, and so on. This visualization shows a two-layer version of a message-passing model since the information is aggregated from two hops. Notice that each node has its computation graph in which the GNN forms a tree structure by unfolding the neighborhood around the target node. The NN modules act as both $\mathcal{G}$ and $\mathcal{F}$ meaning that the input, the aggregated information from node neighbors, passes through to a neural network to generate updated node embedding features.

 Mathematically, for graph $G$, a hidden embedding vector of a node $h_{u}^{k}$ corresponding to each node $u \in V$ is updated according to aggregated information
from $u$’s graph neighborhood $\mathcal{N}(u)$ \cite{hamilton2020graph} 
\begin{align} \label{GNN}
    h_{u}^{(k+1)}= \mathcal{F}^{(k+1)} \left(h_{u}^{(k)}, \mathcal{G}^{k} (\{h_{v}^{(k)}, \forall v \in \mathcal{N}(u)\} ) \right)
\end{align}
where $k$ as iteration ( or layer) indicates number of hope that every node embedding contains information from its k-hop neighborhood. After $k$ iterations, the embedding $h_{u}$ of node $u$ may encode the topological and feature-based information in $k$ hop neighborhood. After running $K$ iterations of the GNN message passing, the output of the final layer can be used to define the embeddings for each node, i.e., $\textbf{z}_{u} = \textbf{h}_{u}(K) , \forall u \in V$.

Depending on the choice of aggregate and update function, there are many GNNs models reviewed in \cite{zhou2020graph}. For example, the basic GNN framework is similar to a multi-layer perception (MLP) as it has linear operations followed by a single element-wise non-linearity. In this paper, Graph Convolutional Network  (GCN) model \cite{kipf2016semi} is considered the SA model. The general idea of GCN is to apply a convolution operator like CNN but over a graph. 

\subsubsection{Graph Convolutional Network}
The GCN is based on spectral methods, in which the representation of a graph lies in the spectral domain, utilizing the Laplacian eigenvectors. The propagation rule is inspired by a first-order approximation of localized spectral filters on graphs.  Given a $N\times M$ feature matrix $X$ ($N$: number of nodes, $M$: number of features), the GCN procedure is as \cite{kipf2016semi}:
\begin{itemize}
    \item  Construing self-connection by adding the identity matrix $I_{N}$ to the adjacency matrix $A$
    \begin{equation} 
    \tilde{A}= A+ I_{N}
    \end{equation}
    \item Using the symmetric normalization of the Laplacian to define convolutional filters
    \begin{equation} 
    L_{norm}= D^{-\frac{1}{2}} L  D^{-\frac{1}{2}}
    \end{equation}
    \item Applying normalization trick to solve exploding/vanishing gradient problems as 
    \begin{equation} 
    I_{N}+D^{-\frac{1}{2}} A D^{-\frac{1}{2}}\xrightarrow{}\Tilde{D}^{-\frac{1}{2}} \tilde{A} \Tilde{D}^{-\frac{1}{2}}
    \end{equation}
    where $\tilde{D}_{ii}= \sum _{j} \tilde{A}_{ij}$ acting as a row-wise summation of the adjacency matrix with self-connection producing the degree of each node.
\end{itemize}

Given $H$ as the feature matrix and $W$ the layer-specific trainable weight matrix, the update rule the layer-wise propagation rule is 
    \begin{equation} 
    H^{(l+1)}= \sigma \left(  \Tilde{D}^{-\frac{1}{2}} \tilde{A} \Tilde{D}^{-\frac{1}{2}} H^{(l)} W^{(l)} \right)
    \end{equation}
    where $\sigma$ is an activation function, such as the $ReLU(\cdot) = max(0,\cdot)$, $H^{(l)}$ $\in \mathbb{R}^{N\times D}$ is the matrix of activations in the $l^{th}$ layer; $H(0) = X$.
After $K$ layer, the GCN produces a node-level output $H^{K}=Z$ (an $N\times F$ feature matrix where $F$ is the
number of output features per node). In this last layer, node embedding $h_{u}^{(K)}$ can pass to a readout layer to present one vector for the final classifier $R$ with learnable parameters to perform graph-level predictions as
\begin{align} \label{readout}
    z_{G}= R \left ( 
    Readout(h_{u}^{(K)}) | u \in G \right)
\end{align}
In fact, the whole procedure lies on the symmetric-normalized aggregation as well as the self-loop update approach for each node as
\begin{align} \label{GCN}
    h_{u}^{(l+1)}= \sigma \left (W^{l} \sum_{v\in \mathcal{N}(u) \cap \{u\}} \frac{h_{u}^{l}}{\sqrt{|\mathcal{N}(u)| |\mathcal{N}(v)| }} \right) 
\end{align}
where $|\cdot| $ indicates the size of node's neighborhoods to train the weight matrix $W$.

Considering the voltage SSA as supervised graph-level classification, the \textit{softmax} function Eq. \ref{softmax} is applied to determine the predicted probability that the graph belongs to the class $G_{i}$.
\begin{align} \label{softmax}
\textit{softmax}(z_{G_i})= \left ( \frac{e^{{z_{G_i}^{T}}}}{\sum_{i=1}^{n} {e^{{z_{G_i}^{T} }}}} \right)
\end{align}

where $z_{G_i}$ is graph-level embedding  over a set of labeled training graphs $T = \{G_{1},\dots, G_{n}\}$.

Therefore, the GCN model is applied for voltage SSA in graph-structured data. The task here is graph-level security classification. Each input is a power grid graph at post-contingency conditions, in which each node has topological and electrical features. The node features are aggregated and updated during the learning procedure with their $k$-neighbors information. Then, last layer classifies the final representations
of each graph obtained by readout layer as secure/insecure. This procedure is repeated to train the weight matrix w.r.t minimizing a loss function. The actual label as secure/insecure is defined according to the reliability criteria of bus voltages in steady-state. 

\section{Simulation Procedure and results}\label{Sec4}
For the simulation purposes, MATLAB and PSS/E software are used for generating different cases. The GCN learning model is implemented using PyTorch Geometric \cite{fey2019fast}.
\subsection{Data Generation} \label{data}
In this paper, the IEEE 118-Bus system is considered for the simulation. This system represents a simple approximation of the American Electric Power system (in the U.S. Midwest) as of December 1962 that contains $19$ generators, $35$ synchronous condensers, $177$ lines, $9$ transformers, and $91$ loads.
The primary practice to generate data in power system applications is to run different contingency scenarios. Due to the highly interconnected nature of modern power systems and various energy market scenarios, operating conditions and even the topology of a power system changes frequently. Capturing all changes in the power system while generating data is an intricate task since the source of variation is unclear, and the power grid operates at various points.
An approach mentioned in Procedure \ref{alg1} is regarded for data generating to provide a rich dataset. Two source of variations are assumed in data generation loop as follows:
\begin{itemize}
    \item \textit{Load variation during day}: The actual net load varies during the day because of the time of day. Here, it is assumed that the 118-bus system follows the same feature as the estimated net load for 2020 from the CAISO "Duck Curve" \cite{wang2017online}. Then, the actual load is scaled by the same factor as the "Duck Curve" for different times of the day. Each scale factor is applied for randomly chosen 70\% of load buses mimicking changes in load profile.
    \item \textit{Stressed conditions}: In this case, the P-V analysis is involved using a series of load flow solutions for incremental power transfers (MW) between \textit{source} (delivers transfer power) and \textit{sink} (absorbs transfer power) areas at the constant power factor. Generators at buses 65, 66, and 69 and load at buses 20, 21, 22, 23, and 115 are considered source and sink areas, respectively. This case results in various operating conditions before the voltages pass the threshold or the load flow does not solve. The P-V solution parameters applied PSS/E setting are 0.5 MW as power mismatch tolerance, and 1000 MW as maximum incremental transfer with an initial transfer of 10 MW by 10 MW transfer increment tolerance.
\end{itemize}
  Voltage operation threshold is assumed $0.90 \text{-}1.10$ p.u. of the steady-state of nominal voltage (post contingency), which is defined by category P1 of TPL-001-WECC-CRT-3.2. The value outside of the range violates the reliability criteria and needs the operators' action.
\newcommand\mycommfont[1]{\footnotesize\ttfamily\textcolor{blue}{#1}}
\SetCommentSty{mycommfont}
\IncMargin{1em}
\begin{algorithm} [ht]
\SetKwData{Left}{left}\SetKwData{This}{this}\SetKwData{Up}{up}
\SetKwFunction{Union}{Union}\SetKwFunction{FindCompress}{FindCompress}
\SetKwInOut{Input}{input}\SetKwInOut{Output}{output}
\tcc{capturing load variation.}
\For{each operating point}{
\For{scenario in the contingency list}{\tcc{ capturing stressed conditions.}
\While{power transfer $<$ maximum incremental transfer\; }{solve power flow\; \eIf {(solution not found {\bf and} voltage violation) }{break, go to the next scenario;}{increase power transfer;}}
}
}
\NoCaptionOfAlgo
\caption{Procedure 1: Data Generation}\label{alg1}
\end{algorithm}\DecMargin{1em}

As a result, $21379$ cases were generated, out of which $19668$ cases were secure, and the remaining $1711$ cases were insecure. The result shows an imbalanced dataset which is rational as the power grid should be secure for most single contingencies ($N \text{-}1$). Each sample indicates a power grid in post-contingency conditions as a graph. Electrical feature obtained by the power flow solution and topological feature computed by centrality measures are then embedded in each node feature vector.
The dataset is randomly split into a 60:20:20 ratio for training, validation, and test sets. The batch size indicating the number of training samples utilized in one training iteration is 128.

\subsection{GCN Model Configuration}
The GCN architecture for voltage SSA as a graph-level classification task is as follows:
\begin{itemize}
    \item \textit{Convolutional layer}: Embedding each node by performing 6 layers of GCN with $\mathrm{ReLU}(x) = \max(x, 0)$ as activation function for each layer, all with a hidden-dimension size of 32.
    \item \textit{Readout layer}: Aggregating node embeddings into a unified graph embedding by averaging the node embeddings
    \item \textit{Final classifier}: A linear classifier with a softmax function to transfer embedding size of hidden-dimension to number of classes.
\end{itemize}
This architecture results 10,946 trainable parameters.
\subsection{Optimization Set up} \label{opt_set}

Considering binary cross-entropy as loss function Eq. \ref{BCEloss}, the model parameters are trained using the adaptive moment estimation (Adam) optimizer with an initial learning rate of $10^{-3}$ and decay the learning rate based on training results to a minimum of $10^{-5}$ for regularization for 200 epochs.
 \begin{align} \label{BCEloss}
 \mathcal{L}= -{(y\log(\hat{y}) + (1 - y)\log(1 - \hat{y}))}
\end{align}
where y is true label and  
$\hat{y}$ denotes the predicted label by Eq. \ref{softmax}.

\subsection{Performance Evaluation Metric} \label{metric}For classification problems, evaluation metrics are used to compare the expected class label to the predicted class label. Since the power system should be operational safe for $(N-1)$ contingency, the majority of the dataset is secure, leading to an imbalanced dataset. Therefore, F1-score and G-mean as the efficient metrics for imbalance data are studied \cite{manning2010introduction} :
\begin{align*}
    \text{F1\text{-}score} = 2 \times \frac{\text{precision} \times \text{recall}}{\text{precision} + \text{recall}} \\
    \text{G\text{-}mean}= \sqrt{\text{recall} \times \text{specificity}}
\end{align*}
Considering confusion matrix in Table \ref{confusion} that indicates all four possible outcomes,we then have 
precision \( =\frac{TP}{TP+FP}\), recall \(=\frac{TP}{TP+FN} \), and specificity \( =\frac{TN}{TN+FP}\).
\begin{table}[t]
\caption{Confusion matrix}
\vspace{-0.2 in}
\label{confusion}
\vskip -0.15in
\begin{center}
\begin{scriptsize}
\begin{sc}
\begin{tabular}{lccccr}
\toprule
 & Predicted   & Predicted \\
  & Positive  & Negative\\
\midrule
Actual & TP & FN\\
Positive & & (Type II error) \\\
Actual& FP &TN\\
Negative& (Type I error) \\
\bottomrule
\end{tabular}
\end{sc}
\end{scriptsize}
\vskip 0.05in
\scriptsize TP: True Positive, FN: False Negative \\ FP: False Positive, TN: True Negative
\end{center}
\vskip -0.1in
\end{table}

\subsection{Case Studies Results}
\begin{figure}[ht]
    \centering
    \includegraphics[scale=0.35]{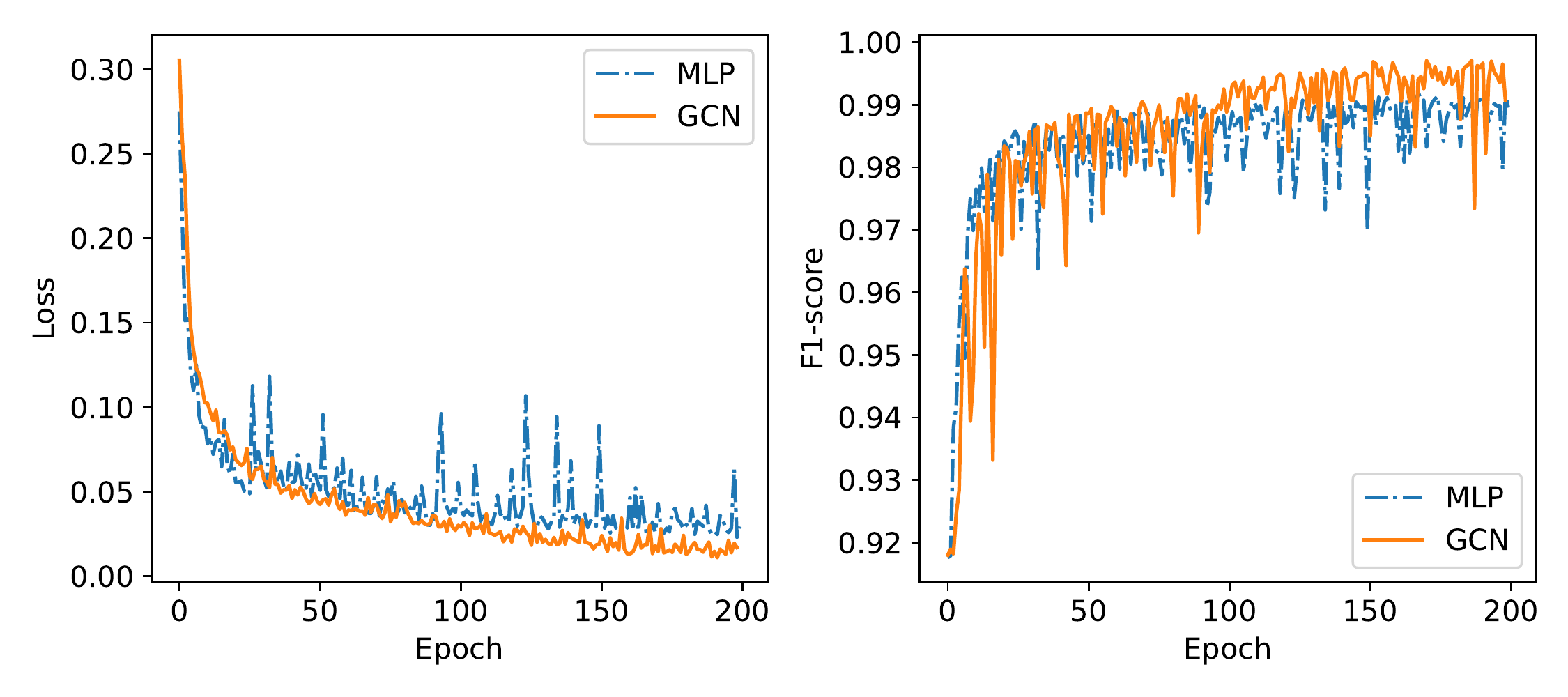}
    \caption{Models F1-score and loss: MLP and GCN comparison}
    \label{fig4}
\end{figure}

As a base case, single line contingency $(N-1)$ scenarios according to Procedure in Section \ref{data} are run to generate samples. The same dataset but in grid-structure data is applied to an MLP model to investigate the application of GCN. For a fair comparison, the MLP model is configured to have a similar number of trainable parameters (10,699) to the GCN model. The MLP model consists of 4 fully connected layers (dense). Other settings, such as the activation function, the number of hidden channels, and optimization parameters, are the same. Training performance metrics and loss are depicted in Fig. \ref{fig4} with electrical and topological input features. As shown, for the same number epoch, not only does GCN provide a better classification result, but its loss value shows convergence to a smaller value than the MLP model. 

Considering the performance metrics mentioned in \ref{metric}, Table \ref{tab1} indicates both GCN and MLP models performance. For investigation on features type's impact on voltage SSA, the models have trained separately for four different groups of input feature
\begin{itemize}
    \item Electrical variables: bus voltage magnitude, bus active and reactive power;
    \item Topological variables: bus degree centrality, bus clustering coefficient, bus betweenness centrality, and bus closeness centrality;
    \item Voltage magnitude plus topological variable;
    \item Both electrical and topological.
\end{itemize}
\begin{table*}[!b]
\vskip 0.15in
\begin{center}
\begin{footnotesize}
\begin{sc}
\caption{Base dataset Performance evaluation: Average value over 10 runs}
\vspace{0.1 in}
\label{tab1}
\begin{tabular}{ccccccccc}
\toprule
&  & &F1-score& & &G-mean& &  \\ \cmidrule(lr){3-5} \cmidrule(lr){6-8}
Variable type & model & Training & Validation  & Test &  Training  & Validation  & Test \\
\midrule
\multirow{2}{*}{Electrical}& GCN & 99.25
&99.08 &99.09 &97.61 &97.04 &97.08& \\
& MLP & 99.06 &99.16 &99.08 &97.02 & 96.30 &96.89 \\
 \midrule
\multirow{2}{*}{Topological} & GCN & 92.43
&91.95 &91.57 &91.30 &91.00&91.23  \\
& MLP & 92.26& 92.29 &91.08 &90.30 &89.05&90.63 \\
\midrule
\multirow{2}{*}{Voltage and topological}   & GCN & 99.36&98.86&99.29 &98.12 &97.22 &97.53& \\
  & MLP & 99.04 &98.32 &99.09 &97.40 &97.02 &97.03& \\
  \midrule
\multirow{2}{*}{Electrical and topological} & GCN & 99.40 &98.74
 &99.50
 & 97.82&96.54 &97.58& \\
  & MLP & 98.87&98.6
 &99.14
 &97.20 &96.26 &96.29& \\  

\bottomrule
\end{tabular}
\end{sc}
\end{footnotesize}
\end{center}
\vskip -0.1in
\end{table*}
The results show that the GCN model outperforms the MLP model as it provides higher evaluation scores. Still considering only topological delivers acceptable performance, and adding the voltage magnitude variable increases the test performance by 8.43\% (F1-score) and 6.90\% (G-mean). This is the expected result, as the voltage variable plays a key role in voltage security assessment. 

Moreover, one may state that there is only a minute improvement in metrics either the GCN model is deployed or topological variables are added. The point here is that improving the performance of an accurate model is challenging since the model may face overfitting. The capacity of the proposed model is analyzed later on. Furthermore, in power system security assessment, the consequences of misclassification may result in a blackout or operation cost. Considering the result of both variables mentioned in Table \ref{tab1}, the GCN model could do the task by providing correct classification by 106 samples (more FP and less FN rate) obtained from F1-score. Regarding practical application, each sample's correct classification is substantial and can be identified via \textit{penalty matrix} \cite{mccalley2010system}. NERC's guidelines approved a matrix comprising violation risk factors and violation severity levels, establishing "base penalty amounts since the protection and remedial action are activated based on classification output. Therefore, slightly enhancing the performance of the SA model not only lowers the risk of the consequences of a wrong decision but also avoids paying penalized fines. 

The other finding refers to the importance of topological variables. Once voltage magnitude is added to topological variables, F1-score and G-mean metrics increase compared to only electrical variables. This indicates topological variables have meaningful intake in the voltage security analysis. The physics behind this observation is not clear. However, the reason would be incorporating the importance of buses across the grids reveals the impact of topological changes. Combining all electrical and topological variables in the feature set provides the highest value of evaluation metrics. In all cases, the GCN model outperforms the MLP because the GCN model captures and updates node information from its neighbors, leading to more informative node features. Indeed, the performance of voltage SSA is enhanced by the usage of information sharing and embedding between buses.
\section{Model Capacity Analysis} \label{Sec5}
To analyze the ability of the built models to adapt properly to new, previously unseen samples so-called \textit{generalization of model}, following scenarios are simulated to observe model capacity.
\subsubsection{Robustness analysis} 
\begin{figure}[ht]
    \centering
    \includegraphics[scale=.5]{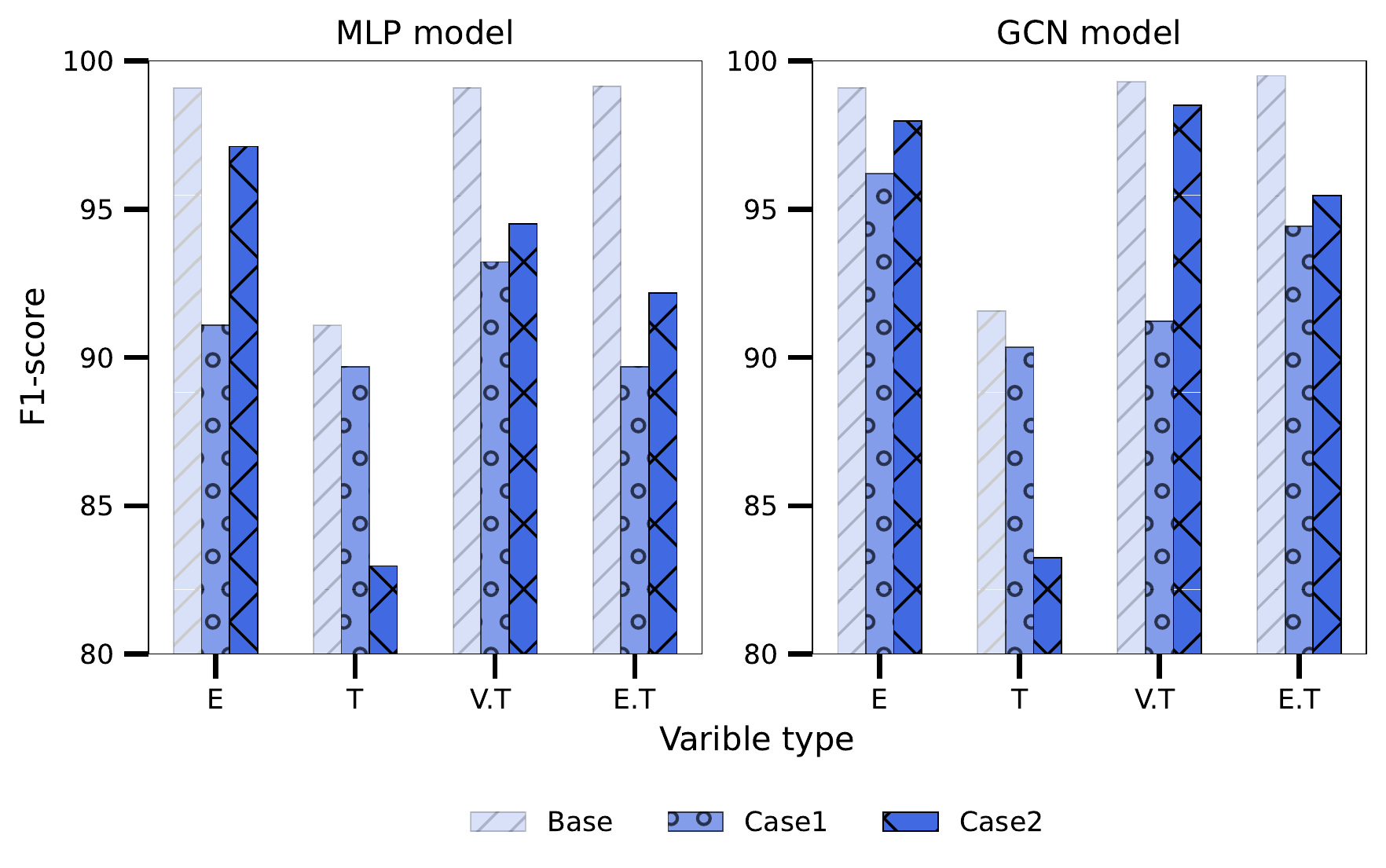}
    \caption{Bar Graph of Robustness Analysis Results}
    \label{comp_all}
\end{figure}

In this case, different datasets are generated to evaluate the robustness of models for unseen samples. For the case 1, a new operating point scaled by "Duck Curve" is generated randomly, and then single line contingencies followed by the increment of power transfer in the system are applied, such as Procedure \ref{alg1}. $1795$ samples are generated in this case. Case 2 for robustness analysis of models is constructed based on double line contingency $(N \text{-}1 \text{-}1)$ from the normal operating point. In this fashion, $15923$ double line scenarios, including base case, are generated without running stressed conditions. Only $500$ cases out of all double line cases are randomly chosen for power transferring for the sake of data generation speed. This results in $5108$ samples. Now, both new datasets are evaluated for the trained models to see the robustness of the SA for unseen datasets. Table \ref{per2} states the robustness analysis results, and performance of models as the bar graph is shown in Fig. \ref{comp_all} for visual observation. As seen, the F1-score drops in both models; however, in both new datasets and for all variable types, the GCN drops less than the MLP model, indicating more model capacity to classify the unseen data. The model's performance varies depending on the variable type. That is, a relation between feature space and new cases is noticed. Since the operating point is associated with the electrical variables, the trained models based on those are more sensitive than double-line contingency cases that address the topological changes. It also observed that trained models based on topological variables are robust for new operating points and sensitive to double-line contingency datasets. Thanks to the combination of both variable types in features, the models perform robustly for both new cases. That is, the model can capture the variation of new datasets from changes in operating point or grid topology. Furthermore, the GCN outperforms the MLP, signifying the functionality of GCN due to feature embedding aggregation and updating operation. \par

\begin{table}
\caption{Robustness Analysis results}

\label{per2}
\vspace{-0.2 in}
\vskip 0.15in
\begin{center}
\begin{footnotesize}
\begin{sc}
\begin{tabular}{cccc}
\toprule
& & \multicolumn{2}{c}{F1-score}\\ \cmidrule(lr){3-4} 
Variable type & model & Case 1& Case 2\\
\midrule
\multirow{2}{*}{Electrical}& GCN & 96.19&97.97\\
  & MLP & 91.09& 97.11 \\
  \midrule
\multirow{2}{*}{Topological}& GCN & 90.34&83.25  \\
& MLP & 89.69&82.97 \\
\midrule
\multirow{2}{*}{Voltage and Topological}   & GCN & 91.22&98.50 \\
  & MLP & 93.22&94.50  \\  
  \midrule
\multirow{2}{*}{Electrical and Topological}    & GCN & 94.42&95.46\\
  & MLP & 89.69&92.17 \\  
\bottomrule
\end{tabular}
\end{sc}
\end{footnotesize}
\vskip 0.05in
\scriptsize Case 1: New operating point, Case 2: Random double line contingency
\end{center}
\vskip -0.1in
\end{table}

\subsubsection{Variables Sensitivity}
Beyond the impact of the unseen dataset, the model can be analyzed when input variables are changed to see how it may impact output classification. Here, sensitivity analysis examines the change in the target output when one of the input features is perturbed. Regarding the state estimation architecture, variables in the SA are estimated from measurement and status, specifically electrical variables. Topological variables might face misinformation generated by the estimation of topology processors due to reported inaccurate data coming from SCADA or limitation of source measurement. However, it is unlikely that changes in the electrical variables. Therefore, variation of voltage magnitude is considered for sensitivity analysis as electrical variables are mainly studied for perturbation analysis due to the device measurement error or being prone to false data injection.

\begin{figure}[ht]
    \centering
    \includegraphics[scale=.65]{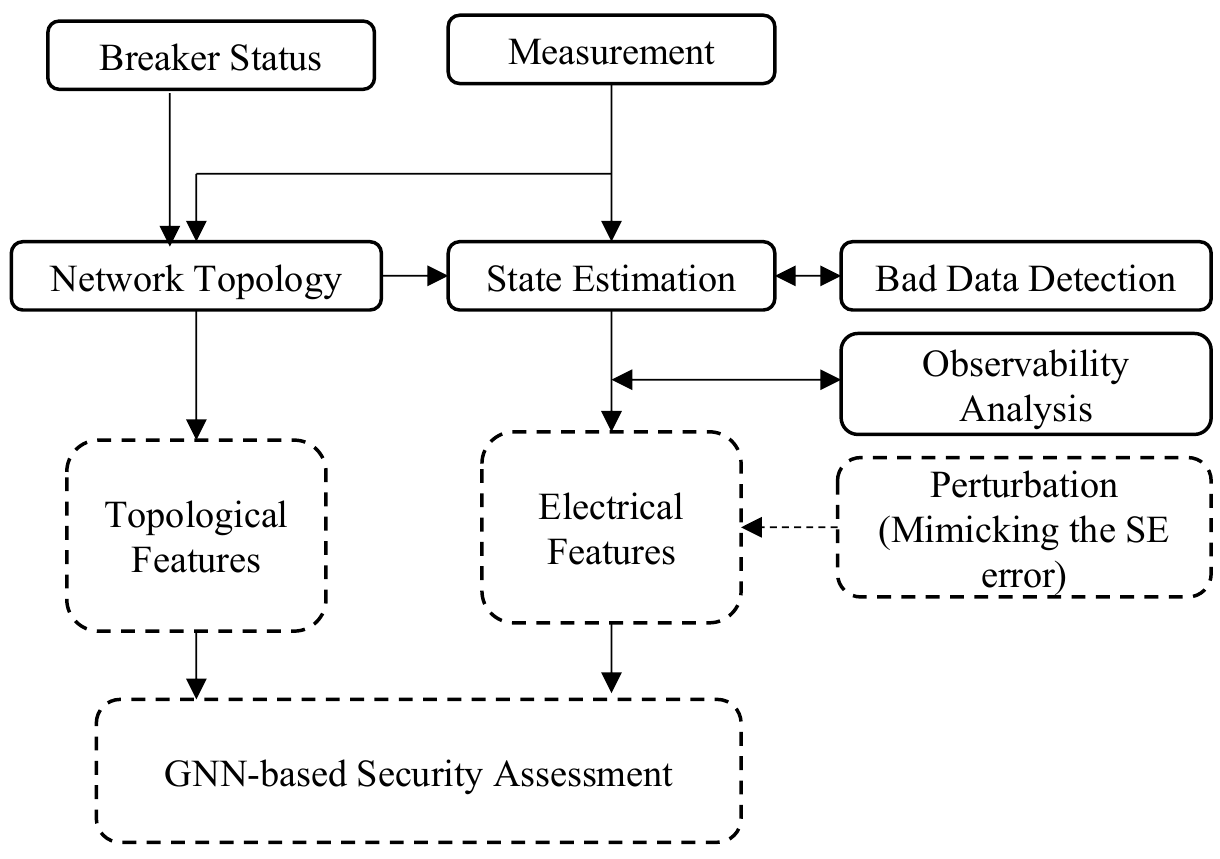}
    \caption{General State Estimation Architecture The dash line blocks are added for sensitivity analysis of GCN based SA model.}
    \label{se}
\end{figure}

Considering a general state estimation architecture shown in Fig. \ref{se} from \cite{abur2004power} related to the SA, a perturbation block is only added to 
to mimic the voltage variation due to the state estimation (SE) error. It should mention that in practice, the disturbances are mainly added to measurements, and here the block only represents the way of voltage variations (SE error) to do sensitivity analysis. We randomly choose 10\% of buses to inject 5\% nominal voltage magnitude as a new test set. One may ask that the SE bad data detection (BDD) scheme can capture this variation and label it as an anomaly before it goes through monitoring and operation. As in practical practice, there is not a specified threshold for BBD, and it also varies depending on operational considerations; the proposed scenario can be applied without practical issues. The models' performance for this scenario is mentioned in Table \ref{per3}, in which the superiority of the GCN model compared to the MLP model is once more seen as the performance of the GCN classifier is less degraded. Although classification performance reduces, using the GCN model with both electrical and topological information results in less sensitivity for voltage magnitude variation. That is clear that this scenario does not apply to models with only topological variables. 

\begin{table}[t]
\caption{Variable Sensitively results}
\vspace{-0.2 in}
\label{per3}
\vskip 0.15in
\begin{center}
\begin{footnotesize}
\begin{sc}
\begin{tabular}{ccc}
\toprule
Variable type & model &F1-score\\
\midrule
\multirow{2}{*}{Electrical}& GCN & 97.90 \\
  & MLP & 94.01  \\
  \midrule
\multirow{2}{*}{Voltage and topological}    & GCN & 98.77  \\
& MLP & 95.66 \\
  \midrule
\multirow{2}{*}{Electrical and topological}    & GCN & 98.86  \\
  & MLP & 95.87 \\

\bottomrule
\end{tabular}
\end{sc}
\end{footnotesize}
\end{center}
\vskip -0.1in
\end{table}

\section{Discussion and Conclusion}\label{Sec6}
This paper introduced the GCN model for voltage SSA by considering the topological variables obtained from the topology of the power grid after contingency scenarios. The following can be discussed in this framework:
\subsection{Do the topological features boost performance?}
As the results showed, the topological variables could enhance the model performance for security assessment and increase the model's capacity through robustness and sensitivity analysis, particularly for the GCN model. Furthermore, the type of centrality measures is also indispensable. Although the topological variables capture the power grid's structure properties, all graph centrality measures do not enhance performance. During the simulation, \textit{harmonic centrality}, which measures the average distance of a node to the other nodes in the network, was also computed. Adding this measure interestingly lowered the best model performance in Table \ref{tab1} by 8.9\%. The reason could be among the topological feature voltage SSA may depend on nodes related measures mentioned in \ref{top_cent}. Therefore, the measures referring directly to distances can be misleading for the voltage SSA. 
\subsection{Training time}
The GNN-based model involves feature aggregation and updates from nodes' neighbors, so the training procedure takes longer than the MLP model. For instance, for both feature types in Table \ref{tab1}, the training time for the GCN and the MLP is 1971 and 63 seconds, respectively. Although the GCN model took much more time for the training (about 32 times the MLP), it resulted in better performance than the MLP for various scenarios. Moreover, the proposed voltage SSA is applied for an offline SA scheme in which training time is not a concern.
\subsection{On a GNN Universal Framework for the SA}
There is no limitation of the proposed GNN-based SA model for voltage SSA to apply for any other security assessments such as frequency, and dynamic security assessment \cite{mukherjee2021real,wang2017online}. The framework is universal such that only the procedure of data generation and reliability criteria are required to be adapted. For example, in voltage DSA of voltage, the dynamic simulation needs to be run, and then bus voltage trajectories must monitor to apply appropriate reliability criteria.

In conclusion, a GCN model as a GNN-based framework is introduced for voltage SSA. Despite the traditional input features such as grid format data, a graph format structure is proposed for security analysis of the power grid. A graph then represents the status of a power grid in which each node (bus) consists of features where the GCN model acts on aggregated and updated bus features from its neighbors to deliver a well-informed design for the SA. In addition to electrical variables, topologically related variables obtained from graph centrality measures indicating a structural property of the power grid are appended to feature space. A dataset capturing possible variations such as daily change of load profile and grid stressed condition for voltage stability are generated for model validation purposes. Simulation results show the outperforming of the GCN model compared to the traditional neural network. The impact of the new feature vector is observed in model performance.
Moreover, the proposed framework is studied for robustness and sensitivity analysis to examine the model's generalization. The outcomes one more confirm the superiority of the GCN model. As the GCN model makes a decision using the neighbor information of each bus where the SA is reformulated in the graph context, it provides more capacity for security classification due to the information distributed over the network. Although this paper focuses on voltage security in the steady state, the developed scheme can simply adapt to any security assessment, such as dynamic, considering other stability problems like frequency.    

\bibliographystyle{IEEEtran}
\bibliography{Ref_GNN}

\end{document}